\documentclass[a4paper]{jpconf}
\usepackage{graphicx}
\usepackage{subfigure}
\usepackage{color}
\usepackage{latexsym}
\begin{document}
\title{Non-equilibrium work distribution for interacting colloidal particles under friction}

\author{Juan Ruben Gomez-Solano$^{1,2}$, Christoph July$^1$, Jakob Mehl$^1$, and Clemens Bechinger$^{1,2}$}

\address{$^1$2. Physikalisches Institut, Universit\"at Stuttgart, Pfaffenwaldring 57, 70569 Stuttgart, Germany}
\address{$^2$Max-Planck-Institute for Intelligent Systems, Heisenbergstrasse 3, 70569 Stuttgart, Germany}

\ead{r.gomez@physik.uni-stuttgart.de}

\begin{abstract}
We experimentally investigate the  non-equilibrium steady-state distribution of the work done by an external force on a mesoscopic system with many coupled degrees of freedom: a colloidal crystal mechanically driven across a commensurate periodic light field. Since this system mimics the spatiotemporal dynamics of a crystalline surface moving on a corrugated substrate, our results show general properties of the work distribution for atomically flat surfaces undergoing friction. We address the role of several parameters which can influence the shape of the work distribution, e.g. the number of particles used to locally probe the properties of the system and the time interval to measure the work. We find that, when tuning the control parameters to induce particle depinning from the substrate, there is an abrupt change of the shape of the work distribution. While in  the completely static and sliding friction regimes the work distribution is Gaussian,  non-Gaussian tails show up due to the spatiotemporal heterogeneity of the particle dynamics during the transition between these two regimes. 
\end{abstract}


\section{Introduction}

\begin{figure}
	\includegraphics[width=1.0\textwidth]{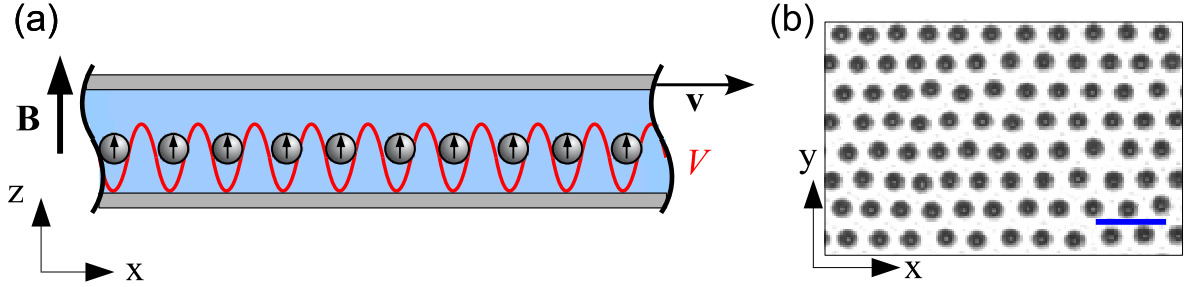}
 	\label{fig:Fig0}
	\caption{(a) Schematic illustration of the sectional view of the experimental setup. (b) Snapshot of the equilibrium crystalline structure of the colloidal monolayer at $B=0.5$~mT, light potential depth $30k_B T$ and $v = 0$. The blue bar represents $20\, \mu\mathrm{m}$. See text for explanation.}
\end{figure}

A basic concept for a system driven in a non-equilibrium process by the change of some external parameters is that 
of work. For mesoscopic systems, e.g. colloidal particles or biomolecules, the work spent in such a process becomes a 
fluctuating quantity which follows a probability distribution of finite width~\cite{sekimoto_a,seifert_a}. During the 
last two  decades, various non-equilibrium work relations, above all the Jarzynski~\cite{jarzynski_a} and the 
Crooks relations~\cite{crooks}, have been established and shown to restrict the shape of this probability distribution 
depending on the underlying specific features of both the system and the non-equilibrium process. From a more 
fundamental point of view, these relations refine the second law of thermodynamics at the mesoscopic scale. Further 
exact statements involving the applied work are rare, even though it has been demonstrated in the framework of 
stochastic thermodynamics~\cite{sekimoto_a,seifert_a,jarzynski_b} that the aforementioned relations as well as 
different ones for other thermodynamic quantities~\cite{evans,gallavotti,kurchan,lebowitz,hatano}, namely entropy 
production and dissipated heat, can be derived from a broader perspective~\cite{seifert_a,seifert_b}.

Experimental tests of non-equilibrium work relations have been carried out for a variety of different systems: for single colloidal particles in 
time-dependent harmonic~\cite{wang,imparato,gomez_a} and non-harmonic potentials~\cite{blickle_a,imparato_b}, 
biomolecules in folding-unfolding assays~\cite{hummer,liphardt,collin,gupta}, mechanical torsion pendulums coupled to 
a heat bath~\cite{douarche}, and charge transitions in electronic devices~\cite{garnier,saira}. Common to all 
experimental systems studied so far is that they only consist of a small number of degrees of freedom, whose internal interactions are irrelevant. From an 
experimental point of view the reason for this is straightforward: controlling all external forces acting on a system 
of many coupled degrees of freedom during a non-equilibrium process in a well-defined way, i.~e., measuring the applied work, is 
a huge challenge~\cite{lander}.

On the other hand, a non-equilibrium phenomenon where the concept of work plays a prominent role is friction. For atomically flat surfaces sliding against each other, friction results from the interplay between externally applied forces and the nonlinear interaction of a large number of degrees of freedom making up extended contacts at the interface. 
Inspired by simplistic models \cite{frenkel}, a 2D system which has attracted much attention in recent years in the field of tribology in order to investigate in a controlled manner  the spatiotemporal dynamics of crystalline surfaces under friction consists of a monolayer of interacting particles suspended in a fluid and mechanically driven through a periodic potential. Since experimental realizations \cite{bohlein} and numerical simulations \cite{vanossi,hasnain} of this system have successfully shed light on the mechanisms behind friction, it represents also an appropriate model to investigate the statistical properties of the non-equilibrium work done by a well-controlled external force on a system composed of many interacting degrees of freedom. So far, this kind of analysis has only been numerically carried out to characterize plastic depinning of interacting particles within a stochastic thermodynamic context \cite{drocco}.

Here, we experimentally study the fluctuations of the work done by an external force on a mesoscopic system with many coupled degrees of freedom: a crystalline monolayer of magnetically interacting colloidal particles moving on a periodic light field under commensurate conditions. The dynamics of this system mimics the transition from static to sliding friction, where a solid surface (the colloidal monolayer) is driven across a corrugated substrate (the periodic light potential) by an external force. 
We investigate the role of several parameters which can influence the shape of the work distribution, e.g. the number of particles used to locally probe the properties of the system and the integration time to measure the work. We find that, when tuning the control parameters to induce particle depinning from the substrate, there is an abrupt change of the shape of the work distribution. While in  the completely static and sliding friction regime, the work distribution is Gaussian,  non-Gaussian tails show up due to the spatiotemporal heterogeneity of the particle dynamics during the transition between these two regimes. Finally, we discuss the asymmetry of the work distribution within the context of the non-equilibrium fluctuation theorems.

\section{Experimental description}

\begin{figure}
	\includegraphics[width=1.0\textwidth]{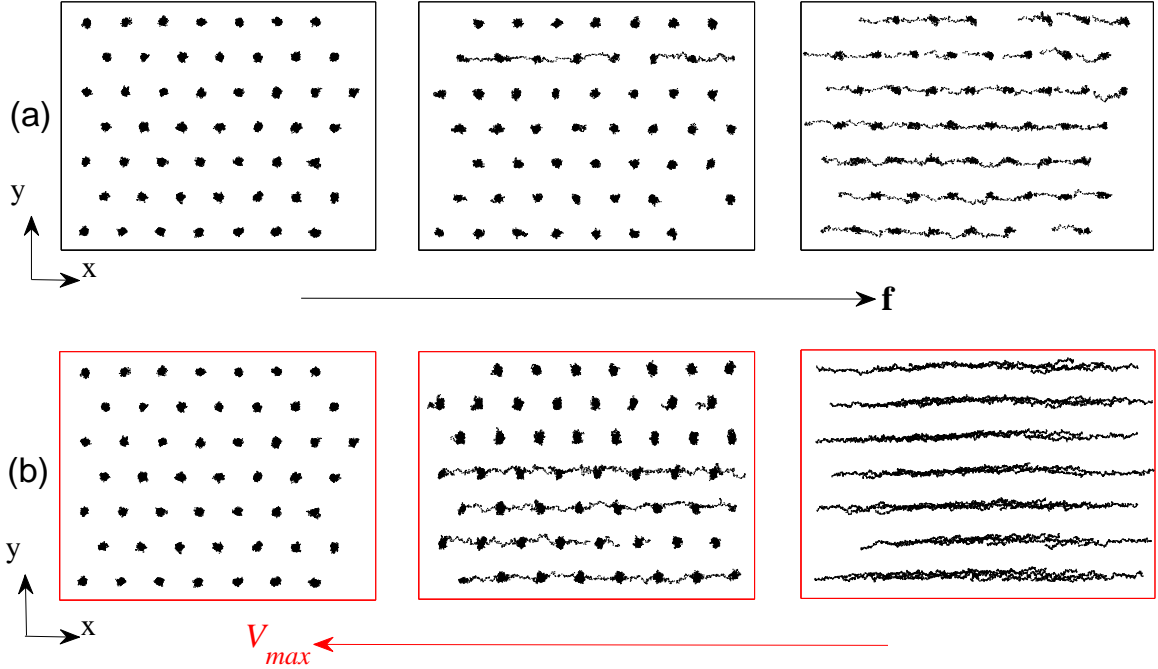}
 	\label{fig:Fig1}
	\caption{(a) Trajectories of 50 neighboring particles measured over  200~s at constant potential depth $V_{max} = 30k_B T$ and different applied forces $f$. From left to right: $f=12.4$~fN, 18.6~fN and 24.8~fN. (b) Trajectories of 50 particles measured over  200~s at constant applied force $f = 12.4$~fN and different potential depths $V_{max}$. From left to right: $V_{max} = 30 k_BT$, $20 k_BT$ and $15k_BT$.}
\end{figure}

Parts of the experimental setup have been described in detail elsewhere~\cite{bohlein} and will be discussed only 
briefly. A schematic illustration of the experimental setup is shown in Fig.~\ref{fig:Fig0}(a). The system consists of a monolayer of superparamagnetic colloidal particles with $2r = 4.5\,\mu\mathrm{m}$ in 
diameter (Dyna M-450 Epoxy, Life Technologies), suspended in a $2.3\,\mathrm{g/l}$ sodium dodecyl sulfate aqueous solution and 
situated in a sample cell of $10\,\mu\mathrm{m}$ height. The total number of particles forming the monolayer is $N \approx 5000$.
Using video microscopy, we simultaneously track the center of mass of approximately 500 particles in the full field of view at 3.3 frames per second with a spatial accuracy of 
$40\,\mathrm{nm}$~\cite{grier}. 
The viscous drag coefficient of the particles in the solvent under this confinement, measured from their equilibrium mean-square displacement, is $\gamma = 6.2 \times 10^{-8}\, \mathrm{kg}\, \mathrm{s}^{-1}$.
A coupling between the particles is obtained by a static homogeneous magnetic field ${\bf{B}} = B{\bf{e}}_z$ applied perpendicular to the sample plane. This field induces a repulsive 
dipole-dipole interaction $U(d)=\mu_0(\chi B)^2/(4\pi d^3)$ with $\mu_0$ the magnetic constant, $d$ the particle separation distance 
and $\chi\simeq3.1\times10^{-11}\mathrm{A}\, \mathrm{ m}^2\,\mathrm{T^{-1}}$ a constant, which allows to control the stiffness of the colloidal 
monolayer. By interference of three laser beams ($\lambda=1064\,\mathrm{nm}$) a light field with hexagonal symmetry inside 
the sample cell is generated, corresponding to a 2D periodic potential landscape $V$, whose maximum depth is $V_{max} = 30{k_B T}$. 
The potential profile is described by the function
\begin{equation}\label{eq:potentialprofile}
	V(x,y) = -\frac{2}{9} V_{max}\left[ \frac{3}{2} + 2 \cos \left(  \frac{2 \pi x}{a} \right)    \cos \left(  \frac{2 \pi y}{\sqrt{3}a} \right) + \cos\left( \frac{4 \pi y}{\sqrt{3}a} \right) \right].
\end{equation}
The lattice constant $a$ and the depth $V_{max}$ can be tuned by the intensities and the angles of incidence of the laser beams~\cite{burns}.  
Before a measurement is performed, the colloidal monolayer is allowed to equilibrate at room temperature $T=298 \pm 0.5$~K for at least one hour in presence 
of a magnetic field of $B=0.5\,\mathrm{mT}$. The resulting homogeneous crystalline state with hexagonal symmetry exhibits a 
lattice constant of approximately $10\,\mu\mathrm{ m}$ with an interaction potential of $U(10\,\mu\mathrm{ m}) \simeq 5 {k_B T}$, as shown in Fig.~\ref{fig:Fig0}(b).
The lattice constant $a$ of the light potential $V$ is adjusted to the same value, i.e. we focus on commensurate conditions, in order to resolve the transition from static to sliding friction of the colloidal monolayer \cite{bohlein}.

\section{Dynamical response under applied force}

\begin{figure}
	\includegraphics[width=1.0\textwidth]{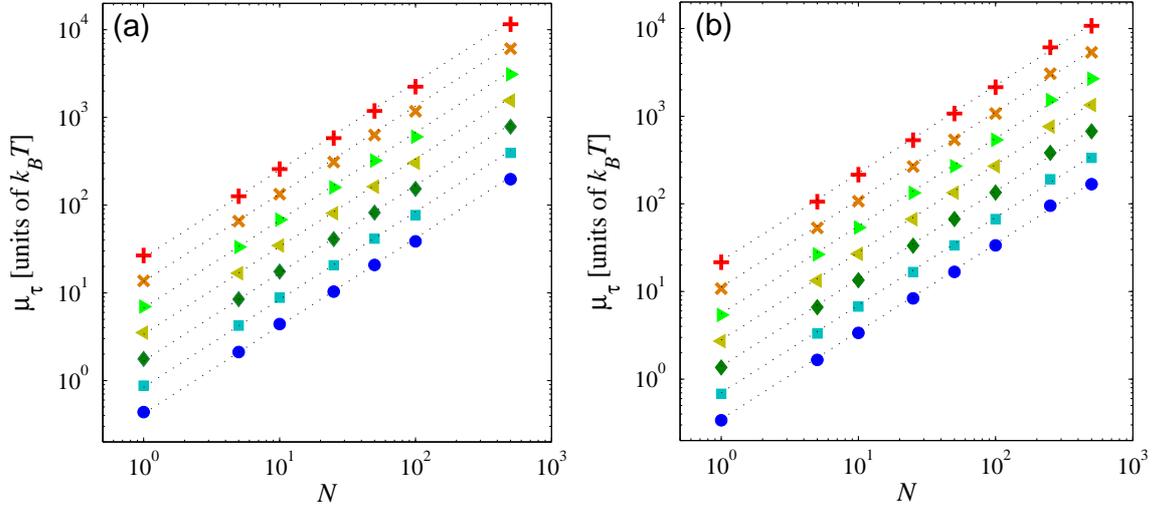}
 	\label{fig:Fign}
	\caption{Mean work done by the external force as a function on the number $N$ of particle trajectories counted in the monolayer for two different frictional regimes (a) $f = 24.8$~fN and $V_{max} = 30k_B T$ (stick-slip motion) and (b) $f = 12.4$~fN and $V_{max} = 15k_B T$ (complete sliding), computed over different time intervals: $\tau = 0.6$~s~($\circ$), 1.2~s~($\Box$), 2.4~s~($\diamond$) , 4.8~s~($\triangleleft$), 9.6~s~($\triangleright$), 19.2~s~($\times$), and 38.4~s~($+$). The dotted lines represent Eq.~(\ref{eq:meanwork}), with no fitting parameters.}
\end{figure}

The whole sample cell is displaced in the x-direction with velocity $v$ by use of a piezo table in order to move the particles across $V$, as sketched in Fig.~\ref{fig:Fig0}(a). The movement allows to create a controlled homogeneous flow ${\bf v}=v{\bf e}_x$ and to drive the 
particles into non-equilibrium steady states (NESS). By tuning the value of $v$ we can mimic the response of the colloidal crystal to the effective external force ${\bf f} \equiv \gamma {\bf v}$. 
The reason of this choice is that many tribological processes take place under these conditions~\cite{carpick,lee,vanossi2}, where an external force is applied to a crystalline surface with the purpose of moving it against a substrate, and where  the surrounding fluid, in our case the solvent, only plays the passive role of a thermal bath to keep the system at fixed temperature.
In order to induce a transition of the monolayer in response to $\bf{f}$ from static to sliding friction, we perform two different experimental protocols. In the first, we keep the depth of the light potential at its maximum value $V_{max} = 30k_B T$ and then we displace the cell at different velocities to tune the magnitude of the applied force $f = |{\bf{f}}|$. In Fig.~\ref{fig:Fig1}(a) we plot some trajectories of 50 neighboring particles moving according to this protocol. At small velocities, corresponding to values of $f$ much smaller than the maximum restoring force exerted by the light potential~(\ref{eq:potentialprofile}), $\max \{ - \nabla V \} = 8\pi V_{max}/(9a)  \approx 34\,\mathrm{fN} $, the particles remained pinned by the potential wells, as shown in the left panel of Fig~\ref{fig:Fig1}(a) for $f = 12.4\,\mathrm{fN}$ ($v = 200\, \mathrm{nm}\,\mathrm{s}^{-1}$). Note that, even when immobile in average, the position of each individual particle fluctuates due to the collision of the solvent molecules. As $f$ approaches values comparable to or larger than $8\pi V_{max}/(9a) $, i.e. when overcoming static friction, the particles are able to move across the potential barriers, thus resulting in collective motion. For example, at $f = 18.6\,\mathrm{fN}$  ($v = 300\, \mathrm{nm}\,\mathrm{s}^{-1}$), the monolayer is partially depinned and some of the particles start to move in the direction of ${\bf{f}}$, as shown in the central panel of Fig.~\ref{fig:Fig1}(a). For $f = 24.8\,\mathrm{fN}$ ($v = 400\, \mathrm{nm}\,\mathrm{s}^{-1}$, right panel of Fig.~\ref{fig:Fig1}(a)), all the particles in the monolayer are able to move in the direction of ${\bf f}$. Under these conditions, the particles undergo stick-slip motion because their mobility is hindered by the presence of the potential barriers~\cite{bohlein,lee}, resulting in a mean particle velocity $105\,\mathrm{nm}\,\mathrm{s}^{-1} < v$.
In the second protocol, we fix the velocity of the sample at $v = 200\, \mathrm{nm}\,\mathrm{s}^{-1}$, i.e. at constant $f = 12.4 \,\mathrm{fN}$, and we vary the depth $V_{max}$ of the light potential. The response of the particles to this protocol with decreasing values of $V_{max}$ is qualitatively similar to that observed when increasing $f$ at constant $V_{max}$ and also has a transition from a static to a sliding friction regime. This effect can be observed in Fig.~\ref{fig:Fig1}(b) for potentials depths $30\,k_BT$, $20\,k_BT$ and $15\,k_BT$, at which the maximum restoring force $8\pi V_{max}/(9a)$ has the values 34~fN, 23~fN and 17~fN, respectively. We point out that, although qualitatively similar, the particle dynamics resulting from these two protocols are not completely equivalent. Indeed, close inspection of the trajectories in Fig.~\ref{fig:Fig1}(a) and ~\ref{fig:Fig1}(b)  shows that in the second protocol, Brownian motion is more significant and the particle mobility is higher because the local confinement created by the substrate is reduced when decreasing $V_{max}$. Therefore, the second protocol reproduces the effect of changing the roughness of the substrate, which in turn results in higher particle velocities compared to the first protocol for the same values of  the parameter $9fa/(8\pi V_{\max})$. For instance, at $9fa/(8\pi V_{\max}) = 0.72$, the mean particle velocity obtained by means of the first protocol is only 26\% of the velocity $v$ of the sample cell (right panel of Fig.~\ref{fig:Fig1}(a)), whereas in the second protocol, it almost reaches free sliding at 93\% of $v$  (right panel of Fig.~\ref{fig:Fig1}(b)).

\section{Stochastic Thermodynamics of the monolayer under applied force}

We focus on the work done on the colloidal monolayer of $N$ interacting particles driven across the corrugation potential $V$ 
by an external force $\bf{f}$, which is the common situation encountered in many tribological problems~\cite{carpick,lee,vanossi2}. We fist present the equations of motion for our specific experimental protocol under flow $\bf{v}$, which allows to mimic in a controlled manner the dynamics under applied force $\bf{f}$. Then, we derive the corresponding stochastic-thermodynamic quantities of the latter tribological process.
When the $i$-th particle ($i=1,...,N$) moves at instantaneous position ${\bf{r}}_i = (x_i,y_i)$ and velocity $\dot{{\bf{r}}}_i = (\dot{x}_i, \dot{y}_i)$ in presence of flow $\bf{v}$, the viscous drag force relative to the flow is $\gamma(\dot{{\bf{r}}}_i - {\bf v})$. In our system, there is no actual external force but only conservative forces derived from the magnetically-induced repulsive interactions and the periodic light field. In addition, each particle is subject to the random thermal collisions of the solvent molecules. Therefore, the dynamics of the $i$-th particle is described by the Langevin equation

\begin{equation}
	\label{eq:langevin}
	\gamma(\dot{{\bf r}}_i -{\bf v}) = -{ \nabla_i}E+{\bf{\xi}}_i,
\end{equation}
where $E$ is the total potential energy of the system, which includes the light potential 
$V$, the pair-interaction potential $U$ of all the particles and the confining potential exerted by the sample cell, $V_{conf}$, which maintains the monolayer in a packed configuration and prevents the particles at the boundaries from escaping from the monolayer due to the repulsive interactions
\begin{equation}
	\label{eq:potential}
	E = \sum_{i=1}^N V({\bf{r}}_i) + \frac{1}{2} \sum_{i = 1}^N \sum_{j \neq i} U(|{\bf{r}}_i - {\bf{r}}_j|) + V_{conf},
\end{equation}
whereas the fast interactions with the surrounding solvent molecules are modeled by a Gaussian white 
noise ${ \xi_i}$ of zero mean and correlations $\langle{ \xi_i}(t){ \xi_j}^T(t')\rangle=2\gamma k_B T\delta(t-t') \delta_{ij}$. 
Because of the structure of Eq.~(\ref{eq:langevin}), the dynamics of every particle in response to the flow and \emph{in absence of an external force}~\cite{mehl} is equivalent to the dynamics in response to an external uniform force, ${\bf{f}} \equiv \gamma {\bf v}$, and \emph{without external flow}\footnote{This equivalence is only valid at sufficiently low Reynolds number, where the flow field around the particle is Stokesian and therefore the resulting drag force can be written as $\gamma(\dot{{\mathrm{r}}}_i - {\bf v})$. In our experiments this assumption is fully justified because the Reynolds number is $\mathrm{Re} < 10^{-4}$.}
\begin{equation}
	\label{eq:langevinforce}
	\gamma\dot{\bf r}_i  = {\bf{f}} -{ \nabla_i}E+{ \xi_i}.
\end{equation}
Therefore, hereafter we will focus only on~Eq.~(\ref{eq:langevinforce}) in order to study the stochastic thermodynamics of a  the monolayer under external constant force and without flow.

In the context of stochastic thermodynamics, the first law for the potential energy variation along a single stochastic realization of the dynamics of the system can be written as \cite{sekimoto_a,sekimoto_b}
\begin{eqnarray}
	\label{eq:firstlaw}
	\mathrm{d}E & = & \sum_{i=1}^N \nabla_i E \cdot \mathrm{d} {\bf{r}}_i, \nonumber\\
	        & = & \mathrm{d}W - \mathrm{d}Q,
\end{eqnarray}
where $\mathrm{d}Q$ and $\mathrm{d}W$ are the heat dissipated into the solvent and the work done by ${\bf{f}}$, respectively, and are given by
\begin{eqnarray}
	\label{eq:workheat}
	\mathrm{d}Q & = & \sum_{i=1}^N [{\bf{f}} - \nabla_i E ] \cdot \mathrm{d} {\bf{r}}_i, \nonumber\\
	\mathrm{d}W    & = &  \sum_{i=1}^N {\bf{f}} \cdot \mathrm{d} {\bf{r}}_i.
\end{eqnarray}
Then, from Eq.~(\ref{eq:workheat}) the work done by the uniform force ${\bf f}$ on the colloidal monolayer, normalized by $k_B T$, over the time interval $[0,\tau]$ reads
\begin{eqnarray}
	\label{eq:workforce}
	w_{\tau}	 & = & \frac{1}{k_B T}\int_0^{\tau}\sum_{i=1}^N  {\bf{f}} \cdot \dot{{\bf{r}}}_i\,\mathrm{d}t \,, 
	\nonumber \\
	 & = &  \frac{f}{k_B T}\sum_{i=1}^N[{x}_i(\tau) -x_i(0) ].
\end{eqnarray}
It should be noted that the expression of the work in Eq.~(\ref{eq:workforce}) only involves the value of the force $f$, which can be tuned experimentally by means of $v$, and the instantaneous values of the x-coordinates of each particle, which are determined by videomicroscopy. Consequently, the work can be directly determined from the particles' trajectories without the need to measure the pair interactions. 
From Eq.~(\ref{eq:workforce}), we can conclude that, regardless of the nature of the pair interactions, the mean value of the NESS work done by $\bf{f}$ over a time interval of duration $\tau$ can be expressed as
\begin{equation}
	\label{eq:meanwork}
	\mu_{\tau} \equiv \langle w_{\tau}\rangle = \frac{Nf\langle \dot{x} \rangle \tau}{k_B T} .
\end{equation}
where the brackets stand for an ensemble average over $N$ particle trajectories and  $\langle \dot{x} \rangle = \frac{1}{N} \sum_{i=1}^N \dot{x}_i$ is the drift particle velocity in response to $\bf{f}$. We check that Eq.~(\ref{eq:meanwork}) is valid in all the frictional regimes investigated in our experiments~\footnote{The linear relation $\mu_{\tau} \propto \tau$ is not necessarily fulfilled for systems with many coupled degrees of freedom driven by time dependent forces, see for example \cite{lacoste}.}. For instance, it is trivially satisfied for static friction, where $\langle \dot{x} \rangle = 0$ yields $\mu_{\tau} = 0$ because in average no mechanical work is done by $\bf{f}$ on the monolayer. On the other hand, for $\langle \dot{x} \rangle > 0$ the linearity of $\mu_{\tau}$ with respect to $N$ and $\tau$ predicted by Eq.~(\ref{eq:meanwork}) is also verified. 
For example, in Figs.~\ref{fig:Fign}(a) and~\ref{fig:Fign}(b) we plot for different integration times $\tau$ the value of the mean work $\mu_{\tau}$ for stick-slip motion and free sliding, respectively, as a function of the number $N$ of particle trajectories used in the computation of $w_{\tau}$. In this case, $\mu_{\tau}$ is determined by taking the average over all possible values of $w_{\tau}$  at fixed $N$ and $\tau$.
We also plot as dotted lines the values of $\mu_{\tau}$ computed by means of Eq.~(\ref{eq:meanwork}), where the NESS drift velocity $\langle \dot{x} \rangle$ is independently determined from the particle dynamics. We observe that the agreement between both kinds of calculations  is excellent.
Therefore, from the validity of Eq.~(\ref{eq:meanwork}) we conclude the mean work mirrors the bulk frictional properties of the monolayer, namely a smooth transition from $\mu_{\tau} = 0$ (static friction with zero mobility at small $f$) to $\mu_{\tau} \propto f^2$ (sliding friction with constant finite mobility at sufficiently large $f$) \cite{bohlein}.

\begin{figure}
	\includegraphics[width=.8\textwidth]{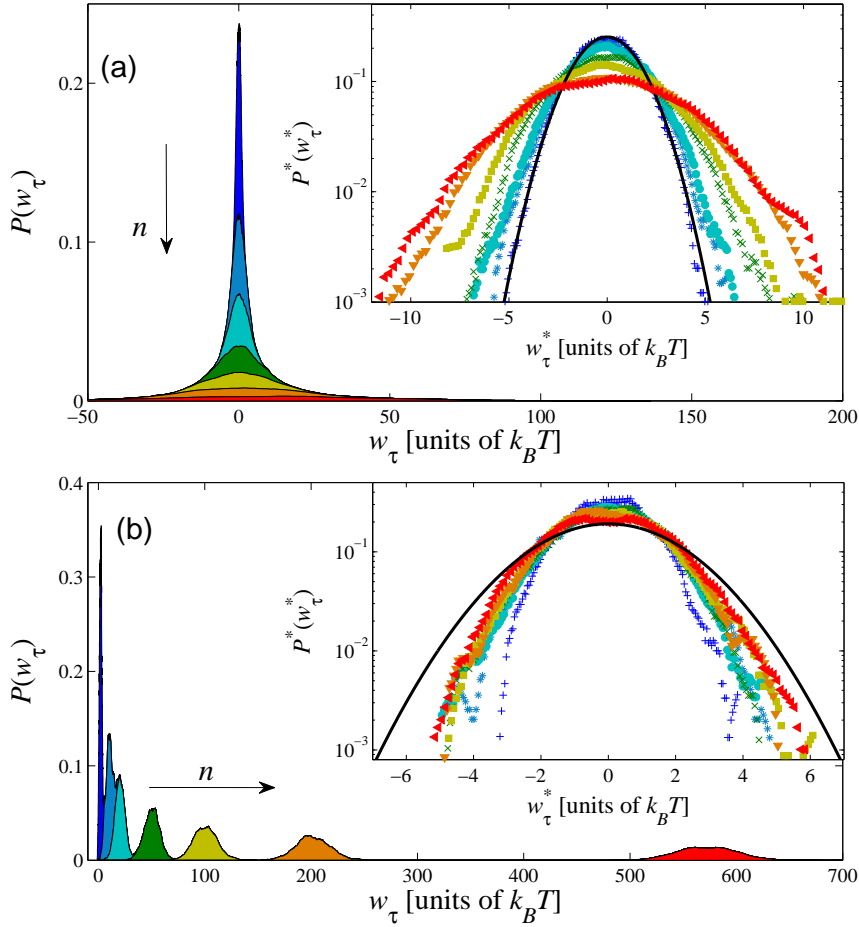}
 	\label{fig:Fig2}
	\caption{Probability density function of the work done by a constant force $f = 12.4$~fN over $\tau=3.6$~s on subsystems composed of different number $n$ of particles across two potentials of depths (a) $V_{max} = 30k_BT$ and (b) $V_{max} = 15k_ BT$.  From top to bottom in~\ref{fig:Fig2}(a) and from left to right in ~\ref{fig:Fig2}(b): $n=1$ (dark blue), 5 (light blue), 10 (cyan), 25 (dark green), 50 (light green), 100 (orange), and 250 (red). Insets: work distribution rescaled according to Eq.~(\ref{eq:rescaling}). The symbols correspond to
 $n=1$ ($+$), 5 ($*$), 10 ($\circ$), 25 ($\times$), 50 ($\Box$), 100 ($\bigtriangledown$), and 250 ($\triangleleft$). The solid lines represent  the rescaled Gaussian distributions for non-interacting particles given by Eqs.~(\ref{eq:case1}) and (\ref{eq:case2}), respectively.}
\end{figure}

\subsection{Non-interacting particles}\label{subsect:Noninteracting}

In principle, the fluctuations of $w_{\tau}$ depend on the strength of the repulsive interactions, the number of particles $N$, the integration time $\tau$, the force ratio $9fa/(8\pi V_{max})$ and the depth $V_{max}$ of the substrate potential. Nevertheless, using the Langevin model of Eq.~(\ref{eq:langevinforce}), we can gain some insight into the statistical properties of the work by analyzing two limit ideal cases which bear resemblance to static and sliding friction, respectively. 

The first case corresponds to $N$ non-interacting particles moving under the influence a very weak force $f \ll 8\pi V_{max}/(9a)$ in presence of a very high potential barrier $V_{max} \gg {k_B T}$, such that the inverse Kramers rate of each particle becomes much larger than the other characteristic time-scales of the system. In this case, which resembles static friction conditions, the system is in a quasi-equilibrium state, where the particles are pinned by the potential wells at an average distance $9fa^2/(16 \pi^2 V_{max})$ from the minima in order to balance the external force $f$.
The probability density function of the work $w_{\tau}$ is Gaussian, i.e. $P(w_{\tau}) = \frac{1}{\sqrt{2\pi\sigma_{\tau}^2}} \exp\left[-\frac{(w_{\tau} - \mu_{\tau})^2}{2\sigma_{\tau}^2}\right]$, with mean $\mu_{\tau}$ and variance $\sigma_{\tau}^2$ given by 
\begin{eqnarray}
	\label{eq:case1}
	\mu_{\tau} & = & 0 \nonumber,\\
	\sigma_{\tau}^2 & = &  \frac{2N f^2 }{ k_B T k} \left[ 1 - \exp\left( -\frac{ k  \tau}{\gamma } \right) \right],
\end{eqnarray}
where $k = [4\pi/(3a)]^2V_{max}$ is the effective stiffness of the restoring force exerted by a periodic light potential with hexagonal symmetry~(\ref{eq:potentialprofile}). Note that, while $\mu_{\tau} = 0$ because no mechanical work is done in average by ${\bf{f}}$, $\sigma_{\tau}^2$ is non-zero. This is due to the thermal fluctuations of the solvent molecules, which can promote either positive or negative work fluctuations by randomly moving the particles with or against the applied force.

The second ideal case is when $N$ non-interacting particles are driven by a sufficiently large force~$f \gg 8\pi V_{max}/(9a)$, such that they move at the highest possible average velocity  $\langle \dot{x} \rangle = f/\gamma$, where the influence of the periodic potential is negligible, similar to free sliding friction. In such a case, the probability density function of $w_{\tau}$ is also Gaussian, with mean and variance
\begin{eqnarray}
	\label{eq:case2}
	\mu_{\tau} & = & \frac{N f^2}{k_B T\gamma} \tau \nonumber,\\
	\sigma_{\tau}^2 & = & \frac{2 N f^2}{k_B T\gamma} \tau = 2 \mu_{\tau},
\end{eqnarray}
respectively. We point out that only in this particular case, the non-equilibrium work trivially satisfies the detailed steady-state fluctuation theorem~\cite{seifert_a}
\begin{equation}
\label{eq:FT}
	\ln \frac{P(w_{\tau} = w)}{P(w_{\tau} = -w)} = \frac{2 \mu_{\tau}}{\sigma_{\tau}^2} w = w,
\end{equation}
because $w_{\tau}$ is actually equal to the total entropy production of the system, normalized by~$k_B$.

In the following, we discuss how the previous ideal expressions for $P(w_{\tau})$ compare to the experimental work distributions for interacting particles in the static and sliding friction regimes. Furthermore, we also investigate the work distribution in the intermediate regime when tuning the control parameters to induce a transition from static to sliding friction of the colloidal monolayer.

\section{Work distribution for interacting particles under applied force}

\begin{figure}
	\includegraphics[width=.75\textwidth]{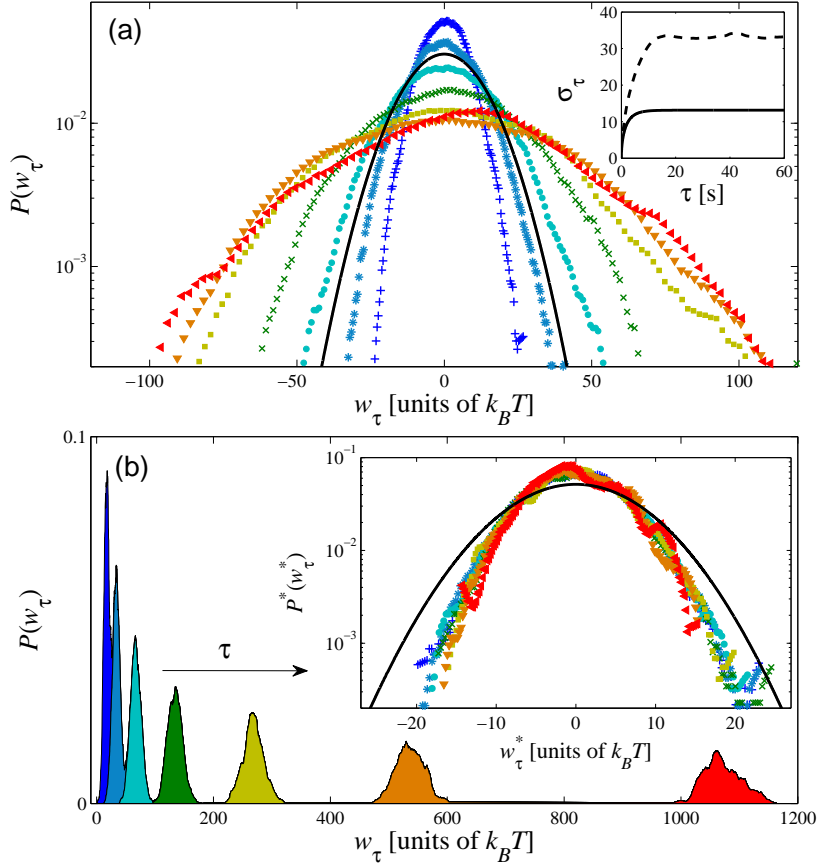}
 	\label{fig:Fig3}
	\caption{(a) Probability density function of the work done by a constant force $f = 12.4$~fN on $n=50$ particles against a  periodic light potential of depth $V_{max} = 30k_BT$ over different time intervals:  $\tau = 0.6$~s~($+$),  1.2~s~($\ast$), 2.4~s~($\bullet$),  4.8~s~($\times$), 9.6~s~($\Box$),  19.2~s~($\bigtriangledown$), and 38.4~s~($\triangleleft$). 
The solid line represents the case of non-interacting particles under the same $V_{\max}$ and $f$. Inset: Standard deviation of $w_{\tau}$ as a function of $\tau$ for non-interacting (solid line)  and magnetically coupled (dashed line) particles (b) Probability density function of the work done by a constant force $f = 12.4$~fN on $n=50$ particles against a light potential of depth $V_{max} = 15 k_BT$ over different time intervals. From left to right:  $\tau = 0.6$~s,  1.2~s, 2.4~s,  4.8~s, 9.6~s,  19.2~s, and 38.4~s. Inset: work distribution measured over different $\tau$ and rescaled according to Eq.~(\ref{eq:rescaling_time}). Same symbols as in Fig.~\ref{fig:Fig3}(a). The solid line represents the case of non-interacting particles.}
\end{figure}

\subsection{Subsystem size}\label{subsect:size}
Since only a portion of the complete monolayer of $N \approx 5000$ particles is accessible for data analysis, a frequent problem encountered in spatially extended systems~\cite{ayton,shang,michel}, we first investigate the effect of measuring the work done on a smaller subsystem composed of $n < N$ particles, thus ignoring its coupling with the $N - n$ degrees of freedom of the rest of the system.  
A possible way to probe the role of such a coupling is by means of the differences between the statistical properties of the work applied on a subsystem of $n$ interacting particles with those observed in a subsystem of the same size $n$ of non-interacting ones, where there is no coupling.
Note that in absence of interactions, the work is a Gaussian variable with mean and variance proportional to the number of components $n$ of the subsystem for the two limit cases described by Eqs.~(\ref{eq:case1}) and (\ref{eq:case2}), i.e. the width of the distribution scales in both cases as $\sigma_{\tau} \propto \sqrt{n}$. 
Therefore, upon translating the work distribution to the origin by an amount $\mu_{\tau}$ and then squeezing it by its width 
\begin{equation}\label{eq:rescaling}
	w^*_{\tau} = \frac{w_{\tau} - \mu_{\tau}}{\sqrt{n}}, \,\,\, P^*(w^*_{\tau}) = \sqrt{n} P(\sqrt{n} w^*_{\tau} + \mu_{\tau})
\end{equation}
any subsystem composed of $n$ non-interacting particles exhibits a $n$-independent profile $ P^*(w^*_{\tau})$.
This means that the statistical properties of the work done on the whole system can be probed by measuring the work done on any subsystem of arbitrary size. 
This situation can change drastically in presence of particle interactions, though. As discussed in~\cite{michel,mehl}, because of the spatio-temporal correlations created by the interactions between the subsystem and the surroundings, the undercount of slow degrees of freedom can give rise to strong modifications of the statistical properties of the thermodynamic quantities of the subsystem with respect to those of the complete system. Then, it is not expected that the variance scales as $\sigma_{\tau}^2 \propto n$ for sufficiently small $n$ in presence of interactions. The effect of the coupling with the surroundings only vanishes when the size of the sampling subsystem spans a length-scale larger than the typical correlation length induced by the interactions, thus recovering the actual statistical properties of the complete system~\cite{michel}.
Indeed, in presence of repulsive interactions we observe this kind of non-trivial dependence of the work distribution on the number $n$ of NESS trajectories used to compute $w_{\tau}$ from Eq.~(\ref{eq:workforce}) in both static and free sliding frictional regimes.

In Fig~\ref{fig:Fig2}(a) we plot the probability density function $P(w_{\tau})$ of the work computed over $\tau = 3.6$~s for subsystems composed of different number $n$ of particles ($n=1,5,10,25,50,100, 250$) at $f = 12.4\,\mathrm{fN}$ and $V_{max} = 30\,k_B T$, for which all the particles in the monolayer are pinned by the potential wells over the observation times accessible  in the experiment, see left panel of Fig.~\ref{fig:Fig1}(a). Each subsystem is chosen in such a way that $n$ neighboring particles cover an approximately square area $\approx n a^2$. We find that for all the values of $n$, $P(w_{\tau})$ is symmetric and peaked around $w_{\tau} = 0$ because no work is done in average in this quasi-equilibrium state, whereas its width increases with increasing $n$. In the inset of Fig.~\ref{fig:Fig2}(a) we plot the work distribution rescaled according to Eq.~(\ref{eq:rescaling}). We observe that $P^*(w_{\tau}^*)$ has a Gaussian profile, and unlike the case of non-interacting pinned  particles, its width increases with increasing $n$. This implies that the variance  $\sigma_{\tau}^2$ grows faster than $n$ in presence of repulsive interactions for sufficiently small $n$, an indication that the correlations between the subsystem and the $N-n$ surrounding particles are significant. Nevertheless,  for sufficiently large values of $n$, we find that $P^*(w_{\tau}^*)$ seems to converge to a size-independent profile, as shown in the inset of Fig.~\ref{fig:Fig2}(a) for $n \ge 100$. The convergence demonstrates that finite-size effects due to the spatial correlations between the sampling subsystem and rest of the particles in the monolayer become negligible compared to the global behavior of $w_{\tau}$ for sufficiently large $n$. However, the effect of the particle interactions on the fluctuations of $w_{\tau}$ persists even for sufficiently large $n$. As a matter of fact, when comparing the experimental $P^*(w_{\tau}^*)$ for $n=100$ and $250$ with that computed from Eq.~(\ref{eq:case1}) with $k=2.2\times10^{-8}\,\mathrm{N}\,\mathrm{m}^{-1}$ for non-interacting particles (solid line in the inset of Fig.~\ref{fig:Fig2}(a)), we find that the former are much wider than the latter. This suggest that, with increasing numbers of particles $n$, the randomness created by the strongly non-linear  coupling accumulate, giving rise to fluctuations of $w_{\tau}$ larger than those that would be otherwise observed in absence of interactions.

A different behavior is observed for free sliding, where all the particles are able to move across the potential landscape at a mean velocity close to $v = f/\gamma$, as those shown in the right panel of Fig.~\ref{fig:Fig1}(b). An example of such a behavior is shown in Fig.~\ref{fig:Fig2}(b) where we plot the probability density function $P(w_{\tau})$ of the work done by a force $f = 12.4$~fN on subsystems formed by different number of particles, $n = 1,5,10,25,50,100, 250$,  across a potential of depth $V_{max} = 15 k_B T$.  In this case, the work distribution is Gaussian, whose maximum is located at positive values of $w_{\tau}$, because the applied force is able to perform  mechanical work by moving the monolayer. The mean work, which coincides with the location of the maximum of $P(w_{\tau})$, increases linearly with increasing $n$, in quantitative agreement with Eq.~(\ref{eq:meanwork}), as shown in Fig.~\ref{fig:Fign}(b). On the other hand, the presence of interactions affects the behavior of the fluctuations of $w_{\tau}$ compared to the ideal sliding case described by Eq.~(\ref{eq:case2}). 
In order to highlight these differences, in the inset of Fig.~\ref{fig:Fig2}(b) we plot the rescaled work distribution $P^*(w^*_{\tau})$ defined in Eq.~(\ref{eq:rescaling}). 
Once more, the effect of the correlation between the subsystem and the rest of the monolayer can be observed for small values of $n$, for which the width of the rescaled distribution increases with $n$. However, for $n > 25$, $P^*(w^*_{\tau})$  converges to a $n$-independent profile, thus probing the actual statistical properties of $w_{\tau}$ for the complete system. 
This convergence implies that the variance of the work scales as $\sigma_{\tau}^2 \propto n$ for sufficiently large $n$.
In the inset of Fig.~\ref{fig:Fig2}(b) we also plot as a solid line the rescaled work distribution of non-interacting sliding particles, described by Eq.~(\ref{eq:case2}). Interestingly, we find that the width of the rescaled work distribution in the presence of interactions is smaller than that of the non-interacting case. We can interpret this narrowing as a reduction of the work fluctuations due to the repulsive interactions, which give rise to an effective stiffening of the monolayer, thus preventing large random excursion of the particles induced by the thermal fluctuations around the drift imposed by $\bf{f}$. This is consistent with the fact that for a perfectly stiff colloidal crystal, which can be realized in the limit of infinitely strong repulsive interactions, thermal fluctuations are  suppressed~\cite{hasnain}, which gives rise to a complete sharpening of the work distribution around the mean value of Eq.~(\ref{eq:meanwork}).

\subsection{Integration time}\label{sect:time}

We now focus on the dependence of the probability density function of the work on the integration time $\tau$. We point out that for values of $\tau$ smaller than the relaxation time-scales of the system,  time-correlations can affect also the statistical properties of the work, because the expression of $w_{\tau}$ in Eq.~(\ref{eq:workforce}) involves differences at distinct times of the particle positions. Nonetheless, for sufficiently large values of $\tau$, such time-correlations vanish and therefore the shape of the work distribution must converge to a single profile upon time rescaling.

We first show in Fig.~\ref{fig:Fig3}(a) the results for the case of a pinned colloidal monolayer, where in average no mechanical work is done. Here we plot the probability density function of the work $w_{\tau}$ done on $n=50$ particles by a force $f = 12.4$~fN against a light potential of depth $V_{max}=30k_B T$ over different integration times, $0.6\,\mathrm{s}\le \tau \le  38.4 \, \mathrm {s}$. We observe that, for all the values of $\tau$. $P(w_{\tau})$ is Gaussian and centered around $w_{\tau}=0$, whose width increases with increasing $\tau$. However, for $\tau > 9.6$~s, the width of the distribution levels off and all curves collapse onto a master curve regarless of $\tau$. This is further verified in the inset of Fig.~\ref{fig:Fig3}(a), where we plot as a dashed line the dependence of the standard deviation $\sigma_{\tau}$ of the work on $\tau$, observing a saturation to a constant value at sufficiently large $\tau$.
The dependence of $\sigma_{\tau}$ on $\tau$ is qualitatively similar to that for non-interacting particles, shown as a solid line in the inset of Fig.~\ref{fig:Fig3}(a). In this case, according to Eq.~(\ref{eq:case1}), the variance of the Gaussian work distribution levels off exponentially for integration times larger than the viscous relaxation time of the particles in the potential wells, $\gamma/k = 2.8$~s. This behavior of $w_{\tau}$ can be actually understood at the single-particle level. For $\tau \ll \gamma/k$, the particle motion is strongly auto-correlated in time due to the energy stored by the confining light potential, which translates into a very narrow distribution $P(w_{\tau})$. The motion becomes less and less correlated when $\tau$ approaches $\gamma/k$, and therefore each particle is able to perform larger Brownian displacements within the potential wells both with and against the applied force, thus resulting in a broadening of $P(w_{\tau})$.
Nevertheless, the fluctuations of $w_{\tau}$ cannot grow indefinitely with increasing $\tau$ because the single-particle motion is always bounded to the potential wells, giving rise to a saturation of $\sigma_{\tau}$ for $\tau >\gamma/k$ .
Although qualitatively similar as a function of $\tau$, we observe a quantitative difference at $\tau \gg  \gamma/k$ between the standard deviation of $w_{\tau}$ for interacting particles with respect to that in the non-interacting case, as shown in the inset Fig.~\ref{fig:Fig3}(a). This difference is due to the strong coupling between the particles, which gives rise to a complex non-linear particle dynamics within the potential wells.

In Fig.~\ref{fig:Fig3}(b) we illustrate the dependence of the work distribution $P(w_\tau)$ on the integration time $\tau$ for the sliding friction regime of $n=50$ particles driven at $f=12.4$~fN and $V_{max}=15k_BT$. 
We find that $P(w_\tau)$ is Gaussian and the location of the maximum increases linearly with increasing $\tau$ in accordance with Eq.~(\ref{eq:meanwork}). Once more, inspired by the comparison with non-interacting particles, where the width of the distribution scales as $\sigma_{\tau} \propto \sqrt{\tau}$ (see Eq.~(\ref{eq:case2})), we can test a scaling with respect to $\tau$ similar to Eq.~(\ref{eq:rescaling})
\begin{equation}\label{eq:rescaling_time}
	w^*_{\tau} = \frac{w_{\tau} - \mu_{\tau}}{\sqrt{\tau}}, \,\,\, P^*(w^*_{\tau}) = \sqrt{\tau} P(\sqrt{\tau} w^*_{\tau} + \mu_{\tau}).
\end{equation}
Interestingly,  in the inset of Fig.~\ref{fig:Fig3}(b) we show that the work distributions, rescaled according to Eq.~(\ref{eq:rescaling_time}), collapse onto a master curve for all $\tau$.
This essentially means that in this frictional regime the variance of the work scales as $\sigma_{\tau}^2 \propto \tau$ even in presence of particle interactions.
Note that in this case, the particles are not confined to move in the potential wells, and consequently there is no intrinsic relaxation time in the dynamics, which explains the very fast convergence of $P^*(w^*_{\tau})$ to the master curve. Quantitative differences are observed between the experimental $P^*(w^*_{\tau})$  and the case without interactions (solid line in the inset of Fig.~\ref{fig:Fig3}(b)), though. This occurs due to the narrowing of the work distribution due to the effective stiffening of the monolayer.

\subsection{Depinning transition}
We now show how the shape of the work distribution changes between the two very distinct cases previously studied, i.e. when changing the experimental parameters to induce a transition from the regime where all the particles are pinned on the substrate, to the depinning of the colloidal monolayer and subsequent free sliding. We recall that in the two extreme regimes of static and sliding friction, the work distribution is Gaussian, even though the mean and variance behave differently as a function of $n$ and $\tau$. 
While in static friction these quantities also depend strongly on both the elastic stiffness $k$ exerted by the substrate potential and the repulsive pair-interactions, they are only affected by the strength of the interactions for sliding friction.

In Fig.~\ref{fig:Fig4}(a) we illustrate the effect on the shape of the work distribution for $n=50$ particles, computed over $\tau=9.6$~s, when increasing the value of the applied force $f$ at constant potential depth $V_{max} = 30k_B T$  in order to induce particle depinning. Interestingly, we observe that the work distribution becomes asymmetric with respect to the maximum with increasing $f$, as can be observed for $f=18.6$~fN and $f=24.8$~fN. In particular, non-Gaussian tails appear at positive values of $w_{\tau}$, as highlighted in the semilog plot of the inset of Fig.~\ref{fig:Fig4}(a). 
For these values of $f$, the spatio-temporal dynamics of the monolayer becomes heterogeneous, as can be observed from the particle trajectories of Fig.~\ref{fig:Fig1}(a). For example, for $f=18.6$~fN there are regions where the particles are still confined by the potential wells, because the external force is still smaller than the maximum conservative force exerted by the light field: $9fa/(8\pi V_{max})=0.54$. However, the combination of thermal fluctuations and non-linear repulsive interactions can promote  hops to the neighboring potential wells, thus creating collective motion of clusters of particles. The collective motion is in turn facilitated by the symmetry beaking induced by the external force. While the stagnant particles do not contribute to the mean value of the work but only to the fluctuations around $w_{\tau} =0$, the non-Gaussian tails originate from the work done on the sliding particles. This regime persists even when the complete monolayer can slide across the periodic substrate, as observed at $f=24.8$~fN and $V_{max} = 30k_B T$, where the particles undergo stick-slip motion. In this case, the spatial heterogeneity is induced by the corrugation potential, creating zones around the potential minima where the particles slow down, whereas they move faster when overcoming the potential barriers, as illustrated by the trajectories in the right panel of Fig.~\ref{fig:Fig1}(a). Note that this heterogeneity results also in a mean particle velocity $\langle \dot{x} \rangle = 105\,\mathrm{nm}\,\mathrm{s}^{-1}$ much smaller than the maximum velocity that could be achieved in presence of a completely flat substrate ($v = 400\, \mathrm{nm}\,\mathrm{s}^{-1}$).

Gaussianity of $w_{\tau}$ is recovered at sufficiently large  $9fa/(8\pi V_{max})$, though, with a respective narrowing of $P(w_{\tau})$. This can be observed in Fig.~\ref{fig:Fig4}(b), where we plot  $P(w_{\tau})$ for the second experimental protocol with which we can reach more easily the free sliding regime. A reduction of only $5k_B T$  in the potential depth $V_{max}$, from $20k_B T$ to $15k_BT$, is enough to observe a prominant change of the shape of the work distribution, as shown in the inset of Fig.~\ref{fig:Fig4}(b). For this values of $V_{max}$, the mean particle velocity changes from $74\,\mathrm{nm}\,\mathrm{s}^{-1}$ to $185\,\mathrm{nm}\,\mathrm{s}^{-1}$, whereas the maximum velocity that could be achieved for this value of $f$ on a completely flat surface is $200\, \mathrm{nm}\,\mathrm{s}^{-1}$.
This dramatic change in the shape of $P(w_{\tau})$  reveals that not only the average tribological properties of the monolayer~\cite{bohlein} but also the properties of the fluctuations of the work done on it become very sensitive when tuning the experimental parameters close to the depinning transition.

\section{Asymmetry of the non-equilibrium work distribution}

\begin{figure}
	\includegraphics[width=.8\textwidth]{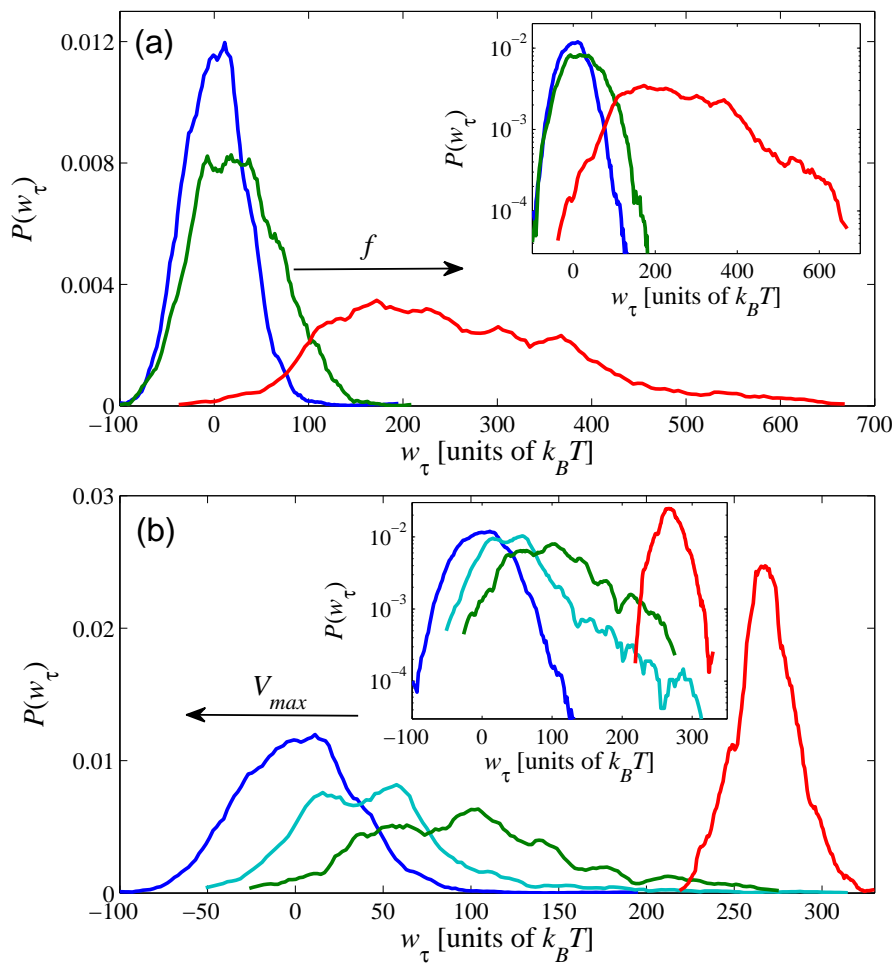}
 	\label{fig:Fig4}
	\caption{(a) Probability density function of the work done by different external forces $f$ on $n=50$ particles against a potential of depth $V_{max}= 30k_BT$, computed over $\tau=9.6$~s. From left to right: $f=12.4$~fN, 18.6~fN, and  24.8~fN. (b)  Probability density function of the work done by an external force $f = 12.4$~fN on $n=50$ particles against potentials of different depth $V_{max}$, computed over $\tau=9.6$~s. From left to right: $V_{max} = 30k_BT,  25k_BT,  20k_BT$, and $15k_BT$. The insets are semi-logarithmic representations of the same plots in the main figures.}
\end{figure}

\begin{figure}
	\includegraphics[width=1.0\textwidth]{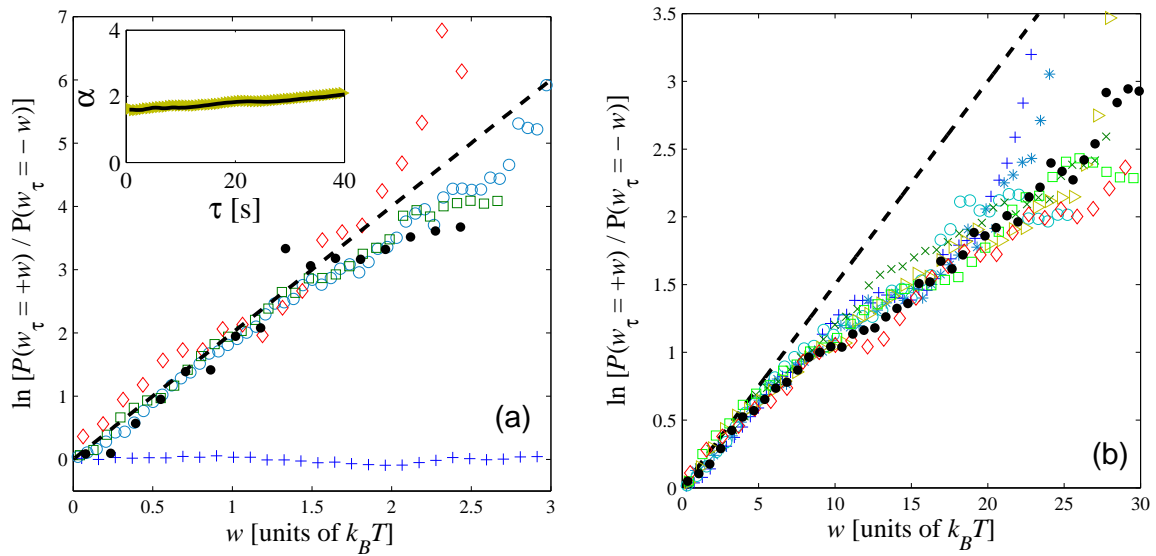}
 	\label{fig:Fig5}
	\caption{(a) Asymmetry function of the work distribution $P(w_{\tau})$ for: static ($+$), and sliding friction regime measured for $n=10$ particles at $f = 12.4$~fN, $V_{max} = 15k_BT$, over $\tau= 0.6$~s ($\circ$), $\tau = 1.2$~s ($\Box$), $\tau = 2.4$~s ($\diamond$). The solid circles are measurements under the same conditions for $n=25$ particles during $\tau = 1.2$~s. The dashed line is a guide to the eye with slope $\alpha = 2$. Inset: dependence of the parameter $\alpha$  on $\tau$ computed by means of Eq.~(\ref{eq:slope}) for $n=100$ ($\triangleright$) and $500$ (solid line) particles.
(b) Asymmetry function of the work distribution $P(w_{\tau})$ measured for $n=25$ particles moving with a heterogeneous dynamics under $f=12.4$~fN and $V_{\max}= 25k_BT$ over $\tau=0.6$~s ($+$), $\tau=1.2$~s ($\ast$), $\tau=2.4$~s ($\circ$), $\tau=4.8$~s ($\times$), $\tau=9.6$~s ($\Box$), $\tau=19.2$~s ($\triangleright$), and $\tau=38.4$~s ($\diamond$).  The solid circles are measurements under the same conditions for $n=50$ particles over $\tau = 38.4$~s. The dashed line is a guide to the eye with slope $\alpha = 0.15$.}
\end{figure}

Finally, we investigate the work distribution $P(w_{\tau})$ within the context of non-equilibrium work relations. More specifically, we focus on the evaluation of quantity $\ln \frac{P(w_{\tau} = +w)}{P(w_{\tau} = - w)}$, which quantifies the asymmetry of the probability of observing positive work fluctuations, where the monolayer moves in the direction of the applied force, with respect to the probability of observing rare negative fluctuations, where the monolayer moves against the force. We point out that, although empirically satisified in many steady-state complex systems \cite{drocco,aumaitre,zamponi,majumdar,joubaud2,hayashi,gradenigo,naert,jimenez,kumar}, a simple linear relation such as Eq.~(\ref{eq:FT}) for the asymmetry function, $\ln \frac{P(w_{\tau} = +w)}{P(w_{\tau} = - w)} \propto w$, is not expected to hold generally for the NESS system we study. Indeed, for this kind of frictional processes, such a linear asymmetry relation is strictly valid  only for the work done by an external force on a collection of non-interacting particles freely sliding on a perfectly flat substrate. In such a case, Eq.~(\ref{eq:FT}) is a direct consequence of the detailed Fluctuation Theorem, which only applies to the total entropy production of the system, and which in that specific case equals the work done by an external force.
On the other hand, since the work defined by Eq.~(\ref{eq:workforce}) has a definite parity under time-reversal, it must satisfy a generalized Fluctuation Theorem in presence of the corrugation potential, the pair-interactions and the global confining potential of the sample cell~\cite{seifert_a}
\begin{equation}
	\label{eq:detailedFT}
	\ln \frac{P(w_{\tau} = +w)}{P(w_{\tau} = - w)} = w- \ln \langle e^{\Delta e_{\tau} - \Delta s_{\tau}}  | w \rangle.
\end{equation}
In Eq.~(\ref{eq:detailedFT}), $\Delta e_{\tau} = [E(\tau) - E(0)]/(k_B T)$ is the variation of the total potential energy of the system during a time interval $\tau$, given by Eq.~(\ref{eq:potential}), $\Delta s_{\tau}$ is the stochastic entropy change over $\tau$~\cite{seifert_b}, and the brackets denote a conditional average over the stochastic realizations for which $w_{\tau}$ equals the value $w$. The last term on the right-hand side of Eq.~(\ref{eq:detailedFT}) is in general non-zero in presence of particle interactions, and therefore $P(w_{\tau})$ does not necessarily satisfy the exact linear relation of Eq.~(\ref{eq:FT}). Note that if $P(w_{\tau})$ is Gaussian, the asymmetry function can still be proportional to $w$, i.e. 
\begin{equation}
	\label{eq:asymmetryGaussian}
		\ln \frac{P(w_{\tau} = +w)}{P(w_{\tau} = - w)} = \alpha w, 
\end{equation}
where the parameter $\alpha$, i.e. the slope of the linear relation, is given by
\begin{equation}
	\label{eq:slope}
		\alpha =\frac{ 2\mu_{\tau} } {\sigma_{\tau}^2}.
\end{equation}
However, unlike the ideal case of non-interacting particle described by Eq.~(\ref{eq:FT}), the parameter $\alpha$ is in general different from 1 because the second term on the right-hand side of Eq.~(\ref{eq:detailedFT}), which involves the particle interactions and the substrate potential, is non-zero.
For instance, we observe that Eq.~(\ref{eq:asymmetryGaussian}) holds in the static friction regime (completely pinned monolayer) and in the sliding regime, as shown in Fig.~\ref{fig:Fig5}(a). In the static friction regime, the parameter $\alpha$ is equal to $0$ for all $n$ and $\tau$, because the system is in a quasi-equilibrium state, with equal probabilities $P(w_{\tau} = +w)$ and $P(w_{\tau} = -w)$. On the other hand, we find  $\alpha \approx 2$ for all $\tau$ in the free sliding regime, as illustrated by the symbols around the dashed line in Fig.~\ref{fig:Fig5}(a). This implies that in this case the second term on the right hand side of Eq.~(\ref{eq:detailedFT}) is non-zero: $\ln \langle e^{\Delta e_{\tau} - \Delta s_{\tau}}  | w \rangle \approx -w$. It should be noted that the direct computation of the asymmetry function from $P(w_{\tau})$ is restricted to rather small values of $n$ and $\tau$, because negative work fluctuations are difficult to sample with increasing values of such parameters. However, taking into account that $P(w_{\tau})$ is Gaussian, $\alpha$ can be estimated from the mean and the variance by means of Eq.~(\ref{eq:slope}). Surprisingly, in the inset of Fig.~\ref{fig:Fig5}(a) we show that the value $\alpha \approx 2$ holds even for $n$ and $\tau$ as large as $500$ and $40$~s, respectively, thus demonstrating that the particle interactions give rise to a robust behavior of the term $\ln \langle e^{\Delta e_{\tau} - \Delta s_{\tau}}  | w \rangle \approx -w$ in Eq.~(\ref{eq:detailedFT}).
This unconventional behavior of the asymmetry function can be traced back to the strong coupling between the particles forming the crystalline monolayer. Indeed, with increasing strength of repulsive interactions, which is fixed in our experiment by the magnetic field $\bf{B}$, the width of the work distribution decreases  because of the increasing stiffness of the colloidal crystal. Note that in the limit of a perfectly stiff colloidal crystal, i.e. created by infinitely large repulsive interactions, the fluctuations of $w_{\tau}$ are completely suppressed. In this case the work distribution becomes a delta function, $P(w_{\tau}) = \delta (w_{\tau} - \mu_{\tau})$,  which gives rise to $\alpha \rightarrow \infty$. Hence, for sliding friction $\alpha$ must be an increasing function of the pair interaction strength,  bounded by the values $\alpha = 1$ (no interactions) and $\infty$ (infinitely large repulsions). The value $\alpha \approx 2$, specific to our experimental conditions, clearly illustrates that repulsive pair interactions reduce the fluctuations of $w_{\tau}$ compared to the value $\alpha= 1$ of Eq.~(\ref{eq:FT}) in absence of interactions.

The intermediate frictional regime, where $P(w_{\tau})$ exhibits non-Gaussian tails due to the heterogeneous spatio-temporal dynamics of the monolayer, is particularly interesting. In this case, we observe that the asymmetry function is not even linear in $w$, as shown in Fig.~\ref{fig:Fig5}(b) for $f=12.4$~fN and $V_{\max}= 25k_BT$, at which only partial depinning from the substrate is achieved. The asymmetry function is approximately linear for small values of $w$, with slope $\alpha \ll 1$ due to large negative work fluctuations on the stagnant particles. Nevertheless, significant deviations from this linear behavior show up at larger work fluctuations, $w>5$, when probing values of $w_{\tau}$ on the non-Gaussian tails plotted in Fig.~\ref{fig:Fig4}(b). Once more, the behavior of the asymmetry functions seems to be robust, as shown in Fig.~\ref{fig:Fig5}(b), where all the data points collapse to a master curve for different values of $\tau$ and $n$.
We point out that in general, such a non-linear behavior of the asymmetry function is not easily observed in systems described by a small number degrees of freedom, because in such a case large negative fluctuations are difficult to sample \cite{mehl}. In our experimental system we are able to achieve this because of the existence of strong negative work fluctuations, which originate from the heterogenous dynamics of  the coupled degrees of freedom of the system during the depinning transition of the colloidal monolayer.

\section{Summary and conclusion}
We have investigated the statistical properties of the work done by an external force on a monolayer of magnetically interacting particles driven across a periodic potential, which mimics friction between crystalline atomic surfaces. We have studied the influence of the number of particles used to probe these properties, the integration time, and the control parameters that are tuned to induce a transition from a pinned state (static friction) to complete depinning from the substrate potential (sliding friction). We have shown that,  in the static and free sliding regimes, the work distribution converges to a Gaussian master curve for sufficiently large number of particles and integration times upon rescaling of these parameters. We have found that the mean and variance of such work distributions depend on the strength of the repulsive interactions, which in particular give rise to a stiffening of the monolayer for free sliding. Interestingly, we have also found that in the intermediate friction regime, where the monolayer undergoes a depinning transition, the work distribution becomes non-Gaussian because of the heterogeneity of the particle dynamics, e.g. due to partial depinning and stick-slip motion. We have shown that in general, the work distribution exhibits unconventional asymmetry properties within the context of non-equilibrium fluctuations relations.  We have demonstrated that such a behavior originates from the presence of repulsive particle interactions. Thus, we provide the first experimental measurements of a stochastic thermodynamic quantity with non-trivial properties for a mesoscopic system with many coupled degrees of freedom.

\section*{Acknowledgments}
We thank Udo Seifert for helpful discussions. We acknowledge financial support of the Deutsche Forschungsgemeinschaft, BE 1788/10-1.

\end{document}